\documentclass[twocolumn,aps,prb,groupedaddress,superscriptaddress]{revtex4-1}
\usepackage[utf8]{inputenc}
\usepackage{amssymb}
\usepackage{verbatim}
\usepackage[usenames]{color}
\usepackage{ulem}
\usepackage{color}
\renewcommand{\emph}[1]{{\it{#1}}}
\usepackage{graphicx}

\begin{document}
\title{Comment on "Pyramidal Structure Formation at the Interface between III/V Semiconductors and Silicon"}

\author{Thomas Hannappel}
\email{thomas.hannappel@tu-ilmenau.de} \affiliation{Technische
Universit\"at Ilmenau, Institut f\"ur Physik, 98693 Ilmenau,
Germany}

\author{Oliver Supplie}
\affiliation{Technische Universit\"at Ilmenau, Institut f\"ur
Physik, 98693 Ilmenau, Germany}

\author{Sebastian Br\"uckner}
\affiliation{Technische Universit\"at Ilmenau, Institut f\"ur
Physik, 98693 Ilmenau, Germany}

\author{Matthias M. May}
\affiliation{Technische Universit\"at Ilmenau, Institut f\"ur
Physik, 98693 Ilmenau, Germany}

\author{Peter Kleinschmidt}
\affiliation{Technische Universit\"at Ilmenau, Institut f\"ur
Physik, 98693 Ilmenau, Germany}

\author{Oleksandr Romanyuk}
\email{romanyuk@fzu.cz} \affiliation{Institute of Physics, Academy
of Sciences of the Czech Republic, Cukrovarnick\'{a} 10, 16200
Prague, Czech Republic}

\begin{abstract}
GaP/Si(100) is considered as pseudomporphic virtual substrate for
III/V-on-Si integration in order to reduce defects related to
polar-on-nonpolar heteroepitaxy. The atomic structure of the
GaP/Si(100) heterointerface is decisive to yield low defect
densities and its dependence on nucleation conditions is still under
debate. Recently, Beyer et al. suggested the formation of a
'pyramidal' structure as a general mechanism at polar-on-nonpolar
interfaces [A. Beyer et al., Chem. Mat. 28, 3265 (2016)]. However,
their DFT studies neglected the dependence of the calculated
interfacial energies on appropriate chemical potentials and their
findings are contradictory to recent and past experimental data.
\end{abstract}

\date{\today}


\maketitle

Recently, Beyer et al.\ reported on 'pyramidal' structure formation
at the GaP/Si(001) heterointerface during low-temperature
nucleation.\cite{BS2016} As a key element of the results, a
'pyramidal' structure formation is suggested to be a general
mechanism at polar-on-nonpolar interfaces,\cite{BS2016} which would
severely impact the interfacial electronic structure and technical
feasibility of nano-dimensioned devices. This generalization,
though, is not justified in the article at any point.\cite{BS2016}
As a second key element, the authors found abrupt Ga-terminated
GaP/Si(100) interfaces to be more stable compared to P-terminated
ones.  These findings were deduced from experimental scanning
transmission microscopy (STEM) high angle annular dark field (HAADF)
images depending on quantitative modeling of the data and
theoretical modeling including DFT calculations. They are in
contradiction to experimental facts that have been published
recently,\cite{SupplieJPCL,SuppliePRB} and also to recent
theoretical calculations.\cite{SuppliePRB,Steinbach} In addition,
the DFT studies in Ref.~[\onlinecite{BS2016}] erroneously neglected
the dependence of the calculated interfacial energies on appropriate
chemical potentials.

The suggested (112) facetted interface model discussed in
Ref.~[\onlinecite{BS2016}] requires an equal number of Si--Ga and
Si--P bonds, which is beneficial with respect to charge neutrality
at the heterointerface. However, applying X-ray photoelectron
spectroscopy, it was clearly shown quantitatively that Si--Ga bonds
do not at all account for 50\,\% of the interfacial bonds (for GaP
nucleation in P-rich conditions).\cite{SupplieJPCL} This
experimental fact is not considered adequately by the authors.
Another experimental observation are low energy electron diffraction
patterns showing the ideal GaP(001)-(2$\times$2)/c(4$\times$2)
surface reconstruction\cite{HahnPRB} after only a few (TBP,TEGa)
pulse pairs--equivalent to ca.\ 3 monolayers of GaP--without any
indications of faceting.\cite{SupplieJPCL}

\begin{figure*}[ht!]
\noindent\centering
\includegraphics[width=11cm]{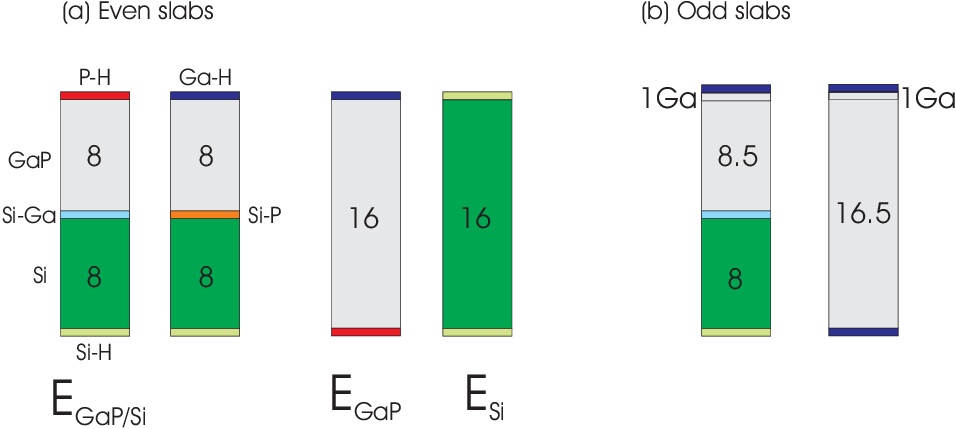}
\caption{Schematic of GaP/Si slabs with (a) even and (b) odd numbers
of layers. In (a), slabs consist of 8 (16) GaP (grey) resp.\ Si
bilayers (green), one P-H (red), one Ga-H (blue) and one Si-H (light
blue) surface, as well as one Si-Ga (cyan) or Si-P (orange)
interface. In (b), the slabs contain an additional Ga
layer.\label{cfig1}}\bigskip
\end{figure*}

With regard to the interfacial structure, it was shown that the
sublattice orientation of the GaP epilayers can be controlled and
inverted in dependence of (i) the amount of Ga available in the
reactor (both with theoretical arguments \cite{SuppliePRB} and
experimental evidence \cite{SupplieACS-AMI}) or also (ii) by a
rotation of the prevailing Si dimer orientation prior to GaP
nucleation.\cite{SuppliePRB} It remains unclear how this could be
achieved with the discussed 'pyramidal' interface structure. Also,
it is not discussed how the step structure at the Si(001) surface
can be preserved during atomic exchange, where Si atoms are assumed
to be replaced across several monolayers, and how the domain
structure of the Si(001) surface prior GaP nucleation corresponds to
the amount of antiphase disorder induced into the GaP epilayer at
the heterointerface. This correlation, however, is an experimental
fact observed also by the very same group.\cite{DoscherAPL}

It should be noted, that the exact Si(001) surfaces discussed in
Ref.~[\onlinecite{BS2016}] are more sensitive to the formation of
elongated vacancy islands on the terraces due to etching of Si by
interaction with the H$_2$ ambient as shown in
Ref.~[\onlinecite{BrucknerNJP}]. Vacancies induce depression on the
terraces in the range of one to few monolayers of Si, which causes
additional anti-phase disorder. The scanning tunneling microscopy
(STM) measurements in Fig.\,S1 in the supplementary material of
Ref.~[\onlinecite{BS2016}] of their Si samples show the presence of
multiple vacancy islands on their terraces and only single atomic
steps. The authors do not comment on the significant occurrence of
these vacancy islands. Their evolution has already been thoroughly
described in Ref.~[\onlinecite{BrucknerNJP}] and could also be an
origin of three-dimensional structures obtained in the STEM analysis
of Ref.~[\onlinecite{BS2016}].

Theoretical investigations of Beyer et.\,al.\cite{BS2016} revealed
that the abrupt Si-Ga interface is energetically more favorable than
the Si-P interface. This conclusion contradicts experimental and
theoretical results in literature
\cite{SupplieJPCL,SuppliePRB,Steinbach}. While relative interface
formation energies were used in Ref.~[\onlinecite{SuppliePRB}],
Beyer et al.\ have computed absolute interface formation energies of
various GaP facets. The interface energy for superlattices was
defined as \cite{BS2016}
\begin{equation}
\Delta E_{if} = [1/2 (E_{Si} + E_{GaP}) - E_{GaP/Si}] / A\quad,
\label{eq1}
\end{equation}
where $E_{Si}$ and $E_{GaP}$ are the total energy of the Si and GaP
slabs, $E_{GaP/Si}$ is the total energy of the slab of the entire
heterostructure,  and  $A$ is the area of the cell (which is assumed
unity for the $(1\times1)$ cell in the following). Superlattices,
however, consist of two interfaces with both polarities and,
therefore, the absolute interface energy of a single interface
cannot be derived by eq.(\ref{eq1}) but only the sum of two
interface energies. The absolute energies were computed from slab
calculations in Ref.~[\onlinecite{BS2016}]. In the following, we
will demonstrate that eq.(\ref{eq1}) is not really valid for slab
calculations.

In Fig.~\ref{cfig1} (a) and (b), slabs with even and odd numbers,
respectively, of Si and GaP layers are shown. The employed slab
configuration consists of GaP layers, Si layers, surfaces and
interfaces with altogether 16 layers (Fig.~6 in
Ref.~[\onlinecite{BS2016}]). The total energies of Si and GaP slabs
have the following contributions:
\begin{eqnarray}
E_{Si} &=& E_{Si}^{bulk} + 2E_{Si-H}\quad,  \label{eq2} \\
E_{GaP} &=& E_{GaP}^{bulk} + E_{Ga-H} + E_{P-H} \quad, \label{eq3}
\end{eqnarray}
where $E_{Si-H}$, $E_{Ga-H}$, and $E_{P-H}$ are the corres\-ponding
surface energies and $E_{Si}^{bulk}$, $E_{GaP}^{bulk}$ are the bulk
energies, i.e.\ the energies of the 16 Si resp.\ GaP bilayers in the
slabs. For the slab with Ga-Si interface, the total energy can be
expressed as
\begin{equation}
E_{GaP/Si} = E_{Si-H} + (E_{Si}^{bulk}+E_{GaP}^{bulk})/2 +
E_{Ga-Si} + E_{P-H}\quad,
 \label{eq4}
\end{equation}
where $E_{Ga-Si}$ is the Ga-Si absolute interface energy. By
substituting eq.(\ref{eq2}), eq.(\ref{eq3}) and eq.(\ref{eq4}) in
eq.(\ref{eq1}), the bulk and the Si-H surface energies are
annihilated---but the surface energies $E_{Ga-H}$ and $E_{P-H}$ are
not. In consequence, the interface energy can be expressed as
\begin{eqnarray}
E_{Ga-Si} &=& \alpha - \Delta E_{if}\quad,  \label{eq6} \\
E_{P-Si} &=& -\alpha - \Delta E_{if}\quad,  \label{eq7} \\
\alpha &=& 1/2 (E_{Ga-H} - E_{P-H})\quad.
\end{eqnarray}
Therefore, the $\Delta E_{if}$ defined in Ref.~[\onlinecite{BS2016}]
is not the absolute interface energy. It involves the surface energy
contributions from two H-terminated surfaces. $\alpha$ depends on
the type of surface reconstruction (termination by H or pseudo H,
for instance) and on the type of surface facet. It is obvious, that
the surface energy of various facets depends on the chemical
potentials $\mu_{Ga}, (\mu_{P})$, and $\mu_{H}$.

Surface energies and $\alpha$ can be derived for polar
semiconductors.\cite{DJ2014,LG2015} It is important, however, that
$\alpha$ depends on the chemical potentials $\mu_{Ga}$, $\mu_{P}$,
and $\mu_{H}$ \cite{DJ2014,HahnPRB} and, therefore, the interface
energy derived from slab calculations with even numbers of
layers,\cite{BS2016} will also depend on the chemical potential.
This is not discussed by Beyer et al.\ at all. The implicit
assumption that $\alpha=0$ for even slabs, i.e.\ $E_{Ga-H} =
E_{P-H}$ for \textit{all facets} of the GaP crystal,\cite{BS2016} is
hard to be rationalized and can lead to misleading conclusions about
the interface stability.

Surface energy contributions can be avoided by using slabs with odd
numbers of layers, i.e. with an additional Ga layer for one type of
the heterostructure [Fig.~\ref{cfig1} (b)]. In this case,
eq.(\ref{eq3}) becomes
\begin{equation}
E_{GaP} = E_{GaP}^{bulk} + 2E_{Ga-H} + E_{Ga} \quad, \label{eq8}
\end{equation}

where $E_{Ga}$ is the energy of additional Ga layer in GaP bulk. It
can be expressed by chemical potentials
\cite{RomanyukPRB88(2013)115312}: In thermodynamic equilibrium, the
chemical potentials are equal to the bulk chemical potentials. In
the particular case discussed here, it holds  $E_{Ga} = \mu_{Ga}$.
By substituting eq.(\ref{eq8}) and eq.(\ref{eq2}) into
eq.(\ref{eq1}), one then obtains the following expressions:
\begin{eqnarray}
E_{Ga-Si} &=& \alpha - \Delta E_{if}\quad,  \label{eq9} \\
E_{P-Si} &=& -\alpha - \Delta E_{if}\quad,  \label{eq10} \\
\alpha &=& -1/2 \mu_{Ga}\quad.
\end{eqnarray}

Similarly to the even case in Fig.\ref{cfig1} (a), the interface
energy depends on the chemical potentials and it is expected to be
different for various facets of the GaP crystal.

Hence, it is important to consider the appropriate chemical
potentials in the DFT calculation of interface energies and, in
particular for even slabs, to involve the surface energy
contributions in order to get correct values.

One reason for the discrepancy in the experimental findings in
Ref.~[\onlinecite{BS2016}] and the observations in literature could
originate in the highly non-equilibrium surface and interface
preparation in metalorganic vapor phase epitaxy, which is sensitive
to smallest disturbances as well as to the history of the growth
reactor. Processes in
Refs.~[\onlinecite{BrucknerPRB,BrucknerNJP,SupplieJPCL,SuppliePRB,SupplieACS-AMI}]
were therefore established with continuous in situ control by
reflection anisotropy spectroscopy.

\bibliographystyle{apsrev}

\end{document}